\documentclass[sigplan,screen]{acmart}

\setcopyright{cc}
\setcctype{by}
\copyrightyear{2025}

\AtBeginDocument{%
  }

\usepackage[toc, acronym]{glossaries}
\usepackage{listings}
\usepackage[T1]{fontenc}
\usepackage[utf8]{inputenc}
\usepackage{import}
\usepackage{float}
\usepackage{natbib}
\usepackage{graphicx}
\usepackage{algorithm}
\usepackage{algpseudocode}
\usepackage{multirow}
\usepackage{xcolor}

\graphicspath{{images}}
\DeclareGraphicsExtensions{.pdf,.jpeg,.png}
\usepackage{amsmath}
\usepackage{url}
\usepackage{booktabs}

\usepackage{microtype}
\usepackage{balance}

\newacronym{rq}{RQ}{research question}
\newacronym{ir}{IR}{intermediate representation}
\newacronym{cpg}{CPG}{code property graph}
\newacronym{ast}{AST}{abstract syntax tree}
\newacronym{cfg}{CFG}{control flow graph}
\newacronym{cdg}{CDG}{control dependence graph}
\newacronym{ddg}{DDG}{data dependence graph}
\newacronym{pdg}{PDG}{program dependence graph}
\newacronym{ssa}{SSA}{static single assignment}

\newacronym{dsl}{DSL}{domain-specific language}
\newacronym{csr}{CSR}{compressed sparse row}
\newacronym{csc}{CSC}{compressed sparse column}
\newacronym{ecs}{ECS}{entity component system}
\newacronym{jmh}{JMH}{Java Microbenchmark Harness}

\newacronym{ifds}{IFDS}{interprocedural finite distributive subset}
\newacronym{ide}{IDE}{interprocedural distributive environment}
\newacronym{lub}{LUB}{least upper bound}
\newacronym{pii}{PII}{personally identifiable information}

\newacronym{llm}{LLM}{large language model}
\newacronym{nlp}{NLP}{natural language processing}
\newacronym{lstm}{LSTM}{long short-term memory}
\newacronym{rnn}{RNN}{recurrent neural network}
\newacronym{gpt}{GPT}{generative pre-trained transformer}
\newacronym{bert}{BERT}{bidirectional encoder representations from transformers}

\newacronym{groum}{GROUM}{graph-based object usage model}
\newacronym{ilp}{ILP}{integer linear programming}
\newacronym{aws}{AWS}{Amazon Web Services}
\newacronym{lcs}{LCS}{longest common subsequence}
\newacronym{sdk}{SDK}{software development kit}

\newacronym{sast}{SAST}{static application security testing}
\newacronym{xss}{XSS}{cross-site scripting}
\newacronym{cfl}{CFL}{context-free language}
\newacronym{api}{API}{application programming interface}
\newacronym{jvm}{JVM}{Java virtual machine}
\newacronym{jar}{JAR}{Java archive}
\newacronym{cicd}{CI/CD}{Continuous Integration/Continuous Deployment}
\newacronym{wpda}{WPDA}{weighted pushdown automata}

\begin{document}

\title{Scalable Language Agnostic Taint Tracking using Explicit Data Dependencies}

\author{Sedick David Baker Effendi}
\orcid{0000-0002-4942-626X}
\affiliation{%
  \institution{Stellenbosch University}
  \city{Stellenbosch}
  \country{South Africa}
}
\email{dbe@sun.ac.za}

\author{Xavier Pinho}
\orcid{0009-0002-8182-5591}
\affiliation{%
  \institution{StackGen}
  \city{San Ramon}
  \country{USA}
}

\author{Andrei Michael Dreyer}
\orcid{0000-0001-5597-5153}
\affiliation{%
  \institution{Whirly Labs}
  \city{Cape Town}
  \country{South Africa}
}

\author{Fabian Yamaguchi}
\orcid{0009-0002-1306-2123}
\affiliation{%
  \institution{Whirly Labs}
  \city{Cape Town}
  \country{South Africa}
}


\begin{abstract}
Taint analysis using explicit whole-program data-dependence graphs is powerful for vulnerability discovery but faces two major challenges. First, accurately modeling taint propagation through calls to external library procedures requires extensive manual annotations, which becomes impractical for large ecosystems. Second, the sheer size of whole-program graph representations leads to serious scalability and performance issues, particularly when quick analysis is needed in continuous development pipelines.

This paper presents the design and implementation of a system for a language-agnostic data-dependence representation. The system accommodates missing annotations describing the behavior of library procedures by over-approximating data flows, allowing annotations to be added later without recalculation. We contribute this data-flow analysis system to the open-source code analysis platform \textsc{Joern}, making it available to the community.
\end{abstract}


\maketitle

\section{Introduction}

Continuous integration and deployment~\citep{fow01} are now standard in many organizations~\citep{gar19}, but achieving continuous vulnerability detection without slowing down the release process remains an ambitious goal. Vulnerability discovery techniques such as symbolic execution~\citep[e.g.][]{kin76,pua09} or fuzz testing~\citep[e.g.][]{bou13,lia18} fall short in this environment, as they assume a relatively static target. Expensive state exploration, however, stands in direct conflict with the need for quick feedback in modern pipelines. 

Researchers have explored using graph databases to store and process whole-program representations of code~\citep{rod20}. In this context, explicit data-dependence representations have proven particularly useful in vulnerability discovery~\citep{kho21}, as they can model a wide range of taint-style vulnerabilities, including command injections, file inclusion vulnerabilities, and \gls*{xss} vulnerabilities. These representations facilitate combining taint propagation information with syntactic and control-flow information to identify vulnerable code~\citep{yam14} and enable automated processing using graph-based machine learning algorithms~\citep{cha21}. However, their accuracy hinges on knowing the taint propagation semantics of all methods.

We present and implement a taint-tracking strategy based on a whole-program data-dependence representation that can be incrementally updated as knowledge about the semantics of external libraries becomes available, avoiding full recomputation when adding new annotations. We contribute our resulting work to an existing open-source code analysis platform, \textsc{Joern}~\citep{joernRepo}, and make it available to the research community. We evaluate the efficiency of our analysis on Java, Python, and JavaScript programs.
\section{Background}
\label{sec:dataflowBackground}

Consider a simple example of data flow that spans external calls to motivate our approach. Listing~\ref{samplecode} defines two methods, \texttt{foo} and \texttt{bar}. The \texttt{foo} method obtains an object \texttt{u} from an external source (\texttt{Source.getValue}), creates a new object \texttt{v}, and calls \texttt{u.transform(v)} to produce \texttt{result}, which is then passed to \texttt{bar(result, v)}. The \texttt{bar} method calls an external method (\texttt{Sink.addValue}) on both its parameters. We want to know whether data from the source, i.e., the return of \texttt{Source.getValue}, can reach the first argument of \texttt{Sink.addValue}. This means checking if a call to \texttt{Source.getValue} \emph{defines} a value $w$ that is later \emph{used} as $w'$ in a call to \texttt{Sink.addValue($w'$)}. In other words, is $w'$ obtained through a series of transformations from $w$?

\noindent 
\begin{minipage}{\linewidth}
\begin{lstlisting}[caption={Sample Java code with methods \texttt{foo} and \texttt{bar} that call external
  methods \texttt{getValue}, \texttt{isPrivileged}, \texttt{addValue}, and \texttt{transform}.}, label={samplecode},language=Java,basicstyle=\footnotesize\ttfamily,captionpos=b]
public class Example {
  public static void foo() {
    Obj u = Source.getValue(); 
    Obj v = new Obj();
    if (Config.isPrivileged()) { 
      Obj result = u.transform(v);
      bar(result, v); // internal
    }
  }
  static void bar(Obj x, Obj y) {
    Sink.addValue(x); // sink
    Sink.addValue(y); // sink 
  }
}
\end{lstlisting}
\end{minipage}

This perspective on code as operations on variables for which arguments are \emph{used}, \emph{defined}, or \emph{used} and \emph{defined} within a method can be expressed via a \emph{data dependence graph}. Originally developed for program slicing~\cite{fer87}, this graph contains edges from nodes describing operations that define a variable to those that use it without prior redefinitions. 

When no information is available about external methods, an analyzer has two options: assume the calls have no effect or assume they taint everything. With the former approach, we obtain an under-approximated graph where no data dependencies are established between the external method calls. With the latter approach, we obtain an over-approximated graph with spurious data dependency paths that may not reflect actual taint propagation.

Compared to the under-approximated graph, the over-approximated graph offers the advantage that the possible data dependency between \texttt{result} at the call to \texttt{bar} and its occurrence at the call to \texttt{Sink.addValue} is indicated by a path in the data dependence graph. Similarly, the potential re-definition of \texttt{u} or \texttt{v} introduced by the call to \texttt{transform} is visible due to the inability to traverse from the node of \texttt{bar} to that of \texttt{Sink.addValue} without passing through that of \texttt{transform}. This is an example of a transitive data dependency, where a dependency from one variable to another is due to a chain of intermediate steps or functions \cite{hor90}.

To deal with these transitive dependencies, \citet{hor90} proposes an elegant extension of intraprocedural data dependence graphs, which they refer to as \emph{system dependence graphs}. In the system dependence graph, separate nodes for input and output arguments are introduced, and transitive dependencies are encoded via direct edges from input to output nodes. This compresses transitive dependencies onto a representation that only perceives local (non-transitive) dependencies.

Nonetheless, the idea of maintaining a representation of data dependencies that is independent of transitive data dependencies -- and that therefore does not need to be recalculated as new information about external methods becomes available -- forms the intellectual basis for our approach. This fits our scenario well, where the behavior of external methods may be characterized in greater detail by the user over time.

\section{Design}
\label{sec:design}

The \textsc{Joern}~\citep{joernRepo} code analysis platform is extended with a data-flow engine. \textsc{Joern}'s language frontends and standard stages generate a unified \gls*{ast}, \gls*{cfg}, and \gls*{cdg}, forming a near-complete \emph{\gls*{cpg}}~\citep{yam14}. The data-flow engine provides the necessary primitives to construct the intermediate \gls*{ddg} for a full \gls*{cpg}. It includes a querying engine to determine flows on the fly for specified sources and sinks. This scheme is a \emph{may} analysis that identifies flows from sources to sinks, considering user-provided semantics of external methods. The following sections describe the design and implementation in greater detail.

\subsection{Data-Dependence Representation}

\begin{figure}[htbp]
  \centering
  \includegraphics[width=0.9\linewidth]{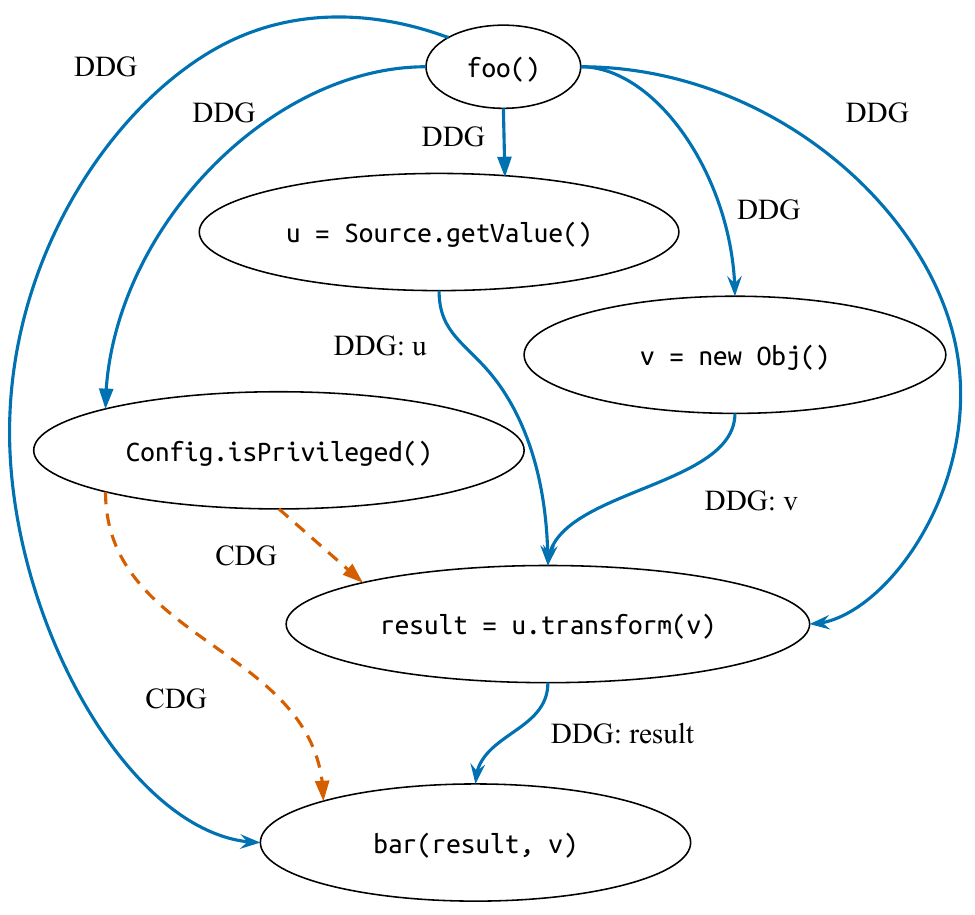}
  \caption{The program dependence graph of the code in Listing~\ref{samplecode}. Edges are labelled as 
    belonging to either the \textcolor[HTML]{D55E00}{control dependence graph (CDG)} or the \textcolor[HTML]{0072B2}{data dependence graph (DDG)}.}
  \label{pdgfig}
  \Description{The graph consists of several nodes, each representing a specific statement in the program, connected by directed edges that indicate dependencies, of which there are two kinds: control dependence edges and data dependence edges. The node labeled ``result = u.transform(v)'' has data dependence edges from ``u = Source.getValue()'' and ``v = new Obj()'', labeled ``u'' and ``v'', respectively. A data dependence edge labeled ``result'' extends from this node to ``bar(result, v)''. The conditional ``Config.isPrivileged()'' introduces a control dependency affecting the execution of both ``result = u.transform(v)'' and ``bar(result, v)'', represented in the graph by a control dependence edge from the former to the latter.}
\end{figure}

The data-dependence representation is based on \glspl*{pdg}~\citep{fer87} (see Figure~\ref{pdgfig}). 

A problem with this approach is that a method’s data-dependence representation is only precise if the semantics of all its transitive callees are known~\citep{hor90}. As \cite{hor90} furthermore shows, summary edges can encode these semantics and be calculated in polynomial time for callee methods available at analysis time. However, for external library methods (with code unavailable at compile time), the user must provide semantics, or they are assumed to be unknown. In this case, it is assumed that all input parameters may taint all output parameters to safely overapproximate the data flow. \citet{palepu2017dynamic} finds success in dynamically generating program summaries for external library code as data and control dependencies between inputs and outputs of calls to external procedures. The authors acknowledge that these summaries can introduce unsoundness and imprecision; however, the performance gains may outweigh these costs. \citet{toman2017taming} explores how library code may bring along many transitive dependencies, and a resulting summary for a method may require referencing indirect flows to other functions. 
The related efforts in summarising external code suggest difficulties in how granular these could be in a language-agnostic approach, and one must accept the inherent imprecision and unsoundness introduced by using such an approach.

To address these challenges, our data-flow engine maintains a stable data-dependence representation as users refine method semantics. It achieves this by treating all callees as external with unknown semantics, over-approximating data dependencies at each call site. Unlike the exploded supergraph in the classical \acrshort*{ifds} framework~\cite{reps95}, this approach does not rely on hard-coded semantics. While this introduces invalid paths, i.e., paths that are not valid in any runtime execution, they are discarded at query time. As a result, adding such summaries is not required to discover additional flows but helps eliminate false positives.
In Figure~\ref{pdgfig}, we note that a call to \texttt{transform} is crucial in determining which paths from the \texttt{Source} to \texttt{Sink} classes are valid. As shown in Listing~\ref{samplesemantics}, we can describe the valid flows for \texttt{transform} to have several outcomes. To define a semantic, one must supply the method's full name, followed by a list of flows between arguments annotated by their positional index and/or argument name, for languages that support named arguments.

Certain positional indexes denote special cases. Such special cases are the return of a call, as index ``\texttt{-1}'', and the receiver as index ``\texttt{0}'', which denotes the object to which the method is bound. Any unspecified flows will be interpreted as \emph{killed} or \emph{sanitized}, i.e., no flow exists between the input and output node. Thus, we need to be explicit where flows are not killed, e.g., \texttt{0->0}. As this may become tedious for methods with many arguments, several special flow objects in the programmatic API provide shorthand ways to define common cases.

To explain how one can define these semantics, we detail the examples from Listing~\ref{samplesemantics}. The first parameter is sanitized (and thus omitted), while the receiver is not and propagates to the return value, defining flows \texttt{0->0} and \texttt{0->-1}. Next, we modify the semantics so that the receiver now taints the non-sanitized first parameter, resulting in flows \texttt{1->1} and \texttt{0->1}.

\noindent 
\begin{minipage}{\linewidth}
\begin{lstlisting}[label={samplesemantics},caption={An example of user-supplied semantics for a call to \texttt{transform}.},moredelim={[s][keywordstyle]{/*}{*/}},basicstyle=\footnotesize\ttfamily,captionpos=b]
/* E.g.1: Argument 1 is sanitized, receiver flow
 propagates to the return value */
"Obj.transform:Obj(Obj)" 0->0 0->-1

/* E.g.2: Receiver taints argument 1 */
"Obj.transform:Obj(Obj)" 0->0 1->1 0->1
\end{lstlisting}
\end{minipage}

Semantics can be written manually or programmatically. One can use heuristics, data-driven approaches, or the data-flow engine to programmatically generate and load new semantics on the fly until one needs to run a data-flow query. 

\subsection{Identifying Data-Flows}
\label{subsec:design-dataflows}

With the data-dependence representation in place, the next step is determining data flows based on user-provided queries. As is true for many other taint analysis systems~\citep[e.g.,][]{bod12, semgrep, codeql}, our query consists of a set of sources and a set of sinks, and it is our goal to determine all source-sink pairs for which a flow from source to sink exists, along with a sample flow. However, as the data-dependence representation does not have hard-coded semantics, a query also includes a set of semantics for external library methods, as we allow the semantics of library methods to change.

Given such a query, the goal now is to calculate data flows in an algorithmically efficient manner that effectively uses multicore CPUs. To this end, an approach similar to ~\citet*{due97} is chosen. They answer queries incrementally, translating queries into tasks and deriving new tasks from the results of prior tasks at method boundaries. Using this approach, each task operates only within the boundaries of a method, such as \texttt{foo} or \texttt{bar} shown in Listing \ref{samplecode}, and can be calculated independently and concurrently. 

Taint analysis can be performed in forward and backwards modes: either traverses data-dependence edges from sources along the edge direction towards sinks or from sinks against the edge direction towards sources.  In the following, only the taint analysis in the backwards direction is described, but the forward direction can be implemented analogously.

A (backwards) task is defined to be given by a start node, an already-known path from the start node to a sink node, the set of source nodes, and the set of semantics. Moreover, a positive integer that indicates analysis depth is stored, referring to the call-chain depth that the analysis explores.

To simplify notation, tasks are described only by pairing start nodes and paths from the start node to the sink; the result table and the analysis depth are assumed to be available. As the sources, sinks, and semantics remain constant throughout the processing of a query, it is assumed that they are available for reading via a globally shared object. With these simplifications in mind, for a given set of sinks $\mathcal{D}$, the initial set of tasks is given by  $\{(d, [], 0) \mid d \in \mathcal{D}\}$, where $[]$ denotes an empty path, and 0 is the initial call depth. 

These tasks are submitted to a work queue, with resulting paths pushed to the output queue. A result can either be \emph{complete}, meaning it describes a flow from a source in $\mathcal{S}$ to a sink in $\mathcal{D}$, or it can be \emph{partial}, meaning that it is a flow that may be part of a complete flow from a source to a sink. We fetch these results from the output queue, record complete results, and derive new tasks from each result. We note that tasks must also be created from complete results, as they may describe sub-flows of a larger complete flow. This procedure is carried out until all tasks have been evaluated and no more new tasks need to be submitted. At this point, all recorded results are returned.

A result is given by a path $p = ([(v_1,r_1), \ldots, (v_N,r_N)], k)$ where $N$ is the path length, and for all $i$ from 1 to $N$, $v_i$ is a node, $r_i$ is a Boolean, and $k$ is the current call depth. For nodes that are arguments in method calls, the Boolean $r_i$ indicates whether the associated method has been resolved in the process of generating the result (\texttt{true}) or whether resolving it has been deferred to a future task (\texttt{false}).

\paragraph{\textbf{Translating results to new tasks.}} From a result $p$, new tasks are generated according to the following rules, shown by Algorithm~\ref{df:alg:translate_tasks}. First, new tasks are only created if the new call depth is no larger than the maximum depth (line 3). If so, do not generate new tasks (line 4), resulting in partial tasks with no new dependent tasks. In this case, flows will be overapproximated for dependent callers of this result. This early termination is a form of widening to ensure termination, analogous to $k$-limiting~\citep{jones1979flow}. Second, if the path begins with a parameter (line 7), we look up the set of corresponding arguments $\mathcal{A}$ (line 8) and generate the tasks $\{ (a,p) \mid a \in \mathcal{A}\}$ (line 9). These corresponding arguments include positional or named arguments at call sites and the receiver of call sites referring to the parameter's method as a higher-order function. Finally, for the given path, all unresolved arguments are determined (line 11). For each unresolved argument, the tasks $\{ (o,p) \mid o \in \mathcal{O} \}$ (line 12) are generated from the set of associated formal output parameters $\mathcal{O}$ (line 11). If the argument is the actual return value of a call, the task $(r, p)$ is also generated, where $r$ denotes the corresponding formal return parameter. If the argument is a method reference, such as a closure, then the closure's return statement becomes a task $(r_c, p)$, where $r_c$ denotes the return statement of the closure.

\begin{algorithm}
    
    \begin{algorithmic}[1]
\Procedure{CreateTasksFromResult}{$p$}
    \State $(x, k) \gets p$ \Comment{Extract the path $x$ and call depth $k$}
    \If{$k + 1 >= k_{\text{max}}$}
        \State \textbf{return} $\emptyset$
    \EndIf
    \State $(v, r) \gets x[0]$ \Comment{Extract head node $v$ and Boolean $r$}
    \If{\Call{IsParameter}{$v$}} 
        \State $\mathcal{A} \gets $ \Call{GetArgsFromCallers}{$v$} 
        \State \textbf{return} $[(a, p, k + 1) \textbf{ for } a \in \mathcal{A}]$ 
    \ElsIf{\Call{IsArgument}{$v$} \textbf{and} $r$ \textbf{is} \textbf{false}}
        \State $\mathcal{O} \gets $ \Call{GetUnresolvedOutArgsAndReturns}{$v$}
        \State \textbf{return} $[(o, p, k + 1) \textbf{ for } o \in \mathcal{O}]$ 
    \Else
        \State \textbf{return} $\emptyset$
    \EndIf
\EndProcedure

    \end{algorithmic}
    \caption{Given a partial result $p$, generates new tasks from parameters and unresolved arguments using the call graph.}
    \label{df:alg:translate_tasks}
\end{algorithm}

\paragraph{\textbf{Solving tasks}} Each task $(s, p, k)$ is solved by a separate worker thread. Results are calculated by inspecting $s$ alone and then determining results for all \emph{valid parents}, that is, nodes with an outgoing data-dependence edge to $s$ that is \emph{valid} according to the semantics $S$.

The result for $s$ is determined as follows. If the head of $p$ is a source, the result is a complete path $(s, \text{false}):p$, where ``$:$'' denotes an append operation. An additional partial path result is pushed if the source is a method parameter. This additional result allows Algorithm~\ref{df:alg:translate_tasks} to create a new task from this result and possibly find additional sources later. If the head of $p$ is not from the source set but a method parameter, then $(s, \text{false}):p$ is returned as the path for a partial result. 

To determine edge validity, the edges from actual returns of method calls are discarded if the semantic value states that the call does not define the return argument. If $s$ is not an argument, we return the remaining list of parents. If $s$ is an argument, incoming edges from parent nodes that are not arguments are valid. In these cases, return a partial result and mark the result as unresolved. These cases either reflect an incomplete call graph or that the task depends on a partial task that was discarded for exceeding the maximum call depth. In either case, the outcome will be that the result is over-approximated, i.e., it is assumed that all of its arguments are both used and defined by a call to the method.

  
\textbf{Validity of parents based on semantics.} A parent node $s_0$ is connected directly to $s$ via an outgoing data-dependence edge, but not all edges are valid according to the semantics. 
For parent nodes that are arguments, if $s$ and $s_0$ are arguments of the same call site and the parent node is used while $s$ is defined according to the semantics, the edge is valid. The edge is also valid if $s$ and $s_0$ are arguments of different call sites, but $s$ is used according to the semantics; otherwise, the edge is invalid. The data flows are over-approximated for methods without defined semantics.

\textbf{Computing results for valid parents.} In the following, we refer to Algorithm~\ref{df:alg:determine_parents}. For each valid parent (lines 3--4), whether a result exists in the table is determined (line 5), and if so, it is used to compute the result by determining the sub-path from the current parent node to the sink and appending $p$, followed by the parent $s_0$. Otherwise, the result is computed recursively; that is, we compute the results for $s_0 : p$. Upon collecting results for all parents and the head node, deduplicate and return (line 13). 

\begin{algorithm}
    \begin{algorithmic}[1]
\Procedure{ComputeResultsForParents}{$s, p, k$}
    \State $\mathcal{R^*} \gets \emptyset$ 
    \ForAll{$s_0 \textbf{ in } $ \Call{Out}{$s$}} \Comment{Traverse DDG edges}
    \If{\Call{IsValidEdge}{$s, s_0$}} 
        \If{$s_0 \in \mathcal{R^*}$} \Comment{Prepend known path}
            \State $\mathcal{R^*} \gets \mathcal{R^*} \cup (\mathcal{R^*}[s_0]:p, k)$ 
        \Else
            \State $r_0 \gets $ \Call{IsOutputArg}{$s_0$}
            \State $\mathcal{R^*} \gets \mathcal{R^*} \cup ((s_0, r_0):p, k)$
        \EndIf
    \EndIf
    \EndFor
    \State \textbf{return} \Call{Deduplicate}{$\mathcal{R^*}$}
\EndProcedure
    \end{algorithmic}
    \caption{Given a task $(s, p, k)$, determine valid results for parents of $s$ using the semantics and data-dependence representation.}
    \label{df:alg:determine_parents}
\end{algorithm}
  
Finally, the union of the results for $p$ and its valid parents is returned. This result is stored in the result table as a cache.

\section{Limitations}

Operators such as assignments, arithmetic, and field accesses are modeled as ordinary call nodes with a default set of semantics. Consequently, aliasing and the heap of data structures are not tracked. In the case of aliases, assignments will propagate flow, but only via weak updates. For data structures that use index accesses for arbitrary keys, such as index values or keys in maps, the data-flow engine tracks these as ``containers'': if an internal member is tainted, then by the semantic definition, the container is tainted.

This leaves future work to make this analysis alias and object-sensitive. However, this imprecision may be attributed to the analysis's low overhead.
\section{Evaluation}
\label{exp1}

This evaluation aims to answer the following research questions: (\textbf{RQ1}) Is the system able to detect taint-style vulnerabilities effectively for multiple programming languages, and (\textbf{RQ2}) without analyzing library code? Finally, an essential property for data-flow analysis in the context of modern programs is (\textbf{RQ3}), i.e., how scalable is our analysis? 

\subsection{Method}
\label{subsec:method}

We compare the precision of the data flow engine of Section~\ref{sec:design} against two static analysis tools that support multiple languages, Semgrep~\citep{semgrep} and CodeQL~\citep{codeql}. The primary considerations for related work are that they are widely adopted, support multi-language taint analysis, and allow partial program analysis. Each tool is run on the same Java, JavaScript, and Python benchmarks, where partial and whole program analysis techniques are compared. We measure precision and recall using the F1-score and Youden's J index~\citep{youden1950index} (rewards higher specificity). We also recorded each tool's analysis runtime and memory usage to assess scalability. All experiments~\cite{baker2025joernBenchmarks} were performed on a platform with a 6-core x86 CPU (3.4 GHz), 32 GB memory, and running Java 21.0.2.

\subsection{Dataset}
\label{subsec:dataset}

Choosing a \emph{well-suited} dataset for taint analysis is non-trivial, where we define well-suited as publicly available and providing a sink, source, and the outcome for any given test.
Securibench Micro~\citep{securibenchMicro} meets these criteria for Java.

While \citet{guarnieri2011saving} mentions developing a benchmark akin to Securibench Micro but for JavaScript, the associated link is dead at the time of writing. To this end, and as a contribution, we develop \emph{securibench-micro.js}~\cite{baker2025securibench} as a JavaScript equivalent to Securibench Micro. For Python, such a benchmark is not readily available; however, an incomplete synthetic benchmark similar in spirit to Securibench Micro exists. As another contribution, this benchmark is completed and dubbed ``Thorat''~\cite{baker2025thorat} after its original author~\citep{thorat}.

When measuring scalability, however, none of the programs in the datasets above compares in magnitude to an industry-sized program. To address this shortcoming, we use Defects4j~\citep{just2014defects4j} and include a Python-inspired variant, namely BugsInPy~\citep{widyasari2020bugsinpy}. While not intended for measuring taint analysis, they include real-world programs that test the scalability of a static analysis tool. The latest versions of each program of these datasets are obtained at the time of writing. 

\begin{table*}[ht]
    \centering 
    \caption{Benchmark results on the partial program static taint analysis for Joern, Joern$_{SEM}$, Semgrep, and CodeQL.}
    \begin{tabular}{lr|ccccccccc}
    \toprule
     Benchmark & Tool & TP & TN & FP & FN & J Index & F1 Score & Runtime (s) & Memory (GB) \\
     \midrule
     \multirow{4}{*}{Securibench Micro} 
        & Joern & 119 & 17 & 36 & 17 & 0.196 & 0.818 & \textbf{1.48$ \pm $0.54} & $0.33 \pm 0.03$ \\
        & Joern$_{SEM}$ & 118 & 36 & 17 & 18 & \textbf{0.547} & \textbf{0.871} & $1.74 \pm 0.59$ & $0.29 \pm 0.02$ \\
        & Semgrep & 100 & 39 & 14 & 36 & 0.471 & 0.800 & $16.73 \pm 0.57$ & \textbf{0.14$ \pm $0.01}\\
        & CodeQL & 93 & 37 & 16 & 43 & 0.382 & 0.759 & $79.57 \pm 1.05$ & $1.26 \pm 0.05$ \\
      \midrule
      \multirow{4}{*}{Thorat}  
       & Joern  & 29 & 22 & 12 & 11 & 0.372 & \textbf{0.716} & $0.91 \pm 0.34$ & $0.26 \pm 0.02$  \\
       & Joern$_{SEM}$  & 29 & 22 & 12 & 11 & 0.372 & \textbf{0.716} & \textbf{0.77$ \pm $0.25} & $0.25 \pm 0.02$ \\
       & Semgrep & 15 & 24 & 10 & 25 & 0.081 & 0.462 & $15.01 \pm 0.41$ & \textbf{0.14$ \pm $0.01} \\
       & CodeQL & 23 & 29 & 5 & 17 & \textbf{0.423} & 0.676 & $44.76 \pm 0.69$ & $0.97 \pm 0.04$ \\
     \midrule
     \multirow{4}{*}{securibench-micro.js} 
      & Joern & 93 & 18 & 26 & 24 & 0.204 & 0.788 & $6.94 \pm 0.49$  & $0.29 \pm 0.02$ \\
      & Joern$_{SEM}$ & 93 & 19 & 25 & 24 & 0.227 & \textbf{0.791} & \textbf{6.88$ \pm $0.47} & $0.28 \pm 0.02$ \\
      & Semgrep & 5 & 43 & 1 & 112 & 0.020 & 0.081 & $14.64 \pm 0.57$ & \textbf{0.14$ \pm $0.01} \\
      & CodeQL & 85 & 31 & 13 & 32 & \textbf{0.431} & \textbf{0.791} & $93.99 \pm 1.98$ & $1.31 \pm 0.05$ \\
    \bottomrule
    \end{tabular}
    \label{tab:results}
\end{table*}

\subsection{Determining a Suitable Analysis Depth}

To justify a suitable value for $k$ to be used by the Joern-based data-flow analysis, one must explore how different values for $k$ affect the results. The results' figures for finding an appropriate value for $k$ have been omitted here for brevity but can be found within the supplementary material.

When measuring against the taint analysis benchmarks, each experiment runs for 10 iterations with user-defined semantics enabled. A significant variation in the J index and F score appears between $k \in [0, 3]$, followed by a slight increase in precision when $k = 8$ in Securibench Micro and securibench-micro.js. When observing the taint analysis wall-clock times for Defects4j and BugsInPy, for a subset of programs, the beginning of exponential complexity for runtime is observed from $k = 6$. Thus, to strike a balance between precision and recall while remaining practical, we determine that $k = 5$ is a safe value for $k$.

\subsection{Taint Analysis}

This section outlines Joern's performance with the presented data-flow engine, using a max call depth $k = 5$, for the three benchmarks against Semgrep and CodeQL.

Table~\ref{tab:results} presents the results of partial taint analysis for each evaluated tool. Joern is assessed in two configurations: without user-defined semantics (Joern) and with manually specified semantics (Joern$_{SEM}$). Both configurations include operator semantics, but the latter incorporates additional, manually curated semantics for external procedure calls, thereby mitigating unnecessary false positives. Whole program analysis is also considered, where the datasets used by Table~\ref{tab:results} are appended with their external dependencies, including transitive ones. The results of whole-program analysis are omitted here but can be found within the supplementary material.

\subsection{Discussion}

Semantic annotations reduce false positives in Securibench Micro, none in Thorat, and one in securibench-micro.js. This is likely due to Java being a more verbose language than Python and JavaScript, leading to the overtainting of more data flows if calls are left unconstrained. Similarly, operations on data structures in JavaScript and Python often use syntactic sugar present as operators, such as index or property accesses.

CodeQL is a reliable choice for analyzing dynamic languages when considering precision, and Semgrep generally falls short. From the evaluation, the Joern-based data-flow analysis can identify the most vulnerabilities; however, this comes at the cost of additional false positives. While the user-defined semantics have been shown to reduce false positives without needing to rerun the analysis, the lack of precision for dynamic languages leaves room for future work.

For whole-program analysis, the Joern results reported fewer false negatives and, in some cases, fewer true positives. The cost of the whole analysis scaled the worst compared to the other candidates. However, it still ended up being the fastest tool in most cases. Compared to the partial-program analysis, a significant cost is incurred for a small precision gain, thus suggesting that the benefits of partial-program analysis outweigh the imprecision.

While memory is not only a direct result of the data-flow analysis, beyond the discrepancy of Joern performing worse in Java, the Joern-based approaches have a memory footprint far closer to that of Semgrep while being closer to CodeQL in precision. However, this figure suggests that CodeQL may scale better than Joern on sufficiently large programs.

Joern's precision for partial program analysis is serviceable and falls somewhere between Semgrep and CodeQL. The individual success of these tools supports the applicability of our work in real-world applications.
The results, interpreted through the constraints of \textbf{RQ1} and \textbf{RQ2}, indicate that our tool effectively and efficiently performs partial-program static taint analysis across multiple programming languages. The results in all categories suggest that the answer to \textbf{RQ3} is that Joern is scalable enough for partial-program analysis of modern programs.
\section{Conclusion}

In response to the growing demand for performant vulnerability discovery in large systems, this paper presented a system capable of language-agnostic static taint analysis without direct access to external dependencies. By using simple annotations to model these dependencies, this system can answer taint analysis queries written with a high-level query language without having to re-analyze the dependencies.

\balance
\bibliographystyle{ACM-Reference-Format}
\bibliography{references}

\onecolumn
\newpage
\appendix
\section{Supplementary Evaluation Figures}
\label{sec:appendix}

\begin{table}[h]
    \centering 
    \caption{Benchmark results on the whole program static taint analysis for Joern, Semgrep, and CodeQL. As Semgrep does not parse bytecode, the Java results are omitted.}
    \begin{tabular}{lr|ccccccccc}
    \toprule
      Benchmark & Tool & TP & TN & FP & FN & J Index & F1 Score & Runtime (s) & Memory (GB) \\
     \midrule
     \multirow{3}{*}{Securibench Micro} 
        & Joern & 113 & 37 & 16 & 23 & \textbf{0.529} & \textbf{0.853} & \textbf{103.27$ \pm $5.50} & $1.91 \pm 0.03$ \\
        & Semgrep & - & - & - & - & - & - & - & - \\
        & CodeQL & 93 & 37 & 16 & 43 & 0.382 & 0.759 & 109.73$ \pm $2.20 & \textbf{1.53$ \pm $0.03} \\
      \midrule
      \multirow{3}{*}{Thorat}  
       & Joern  & 27 & 23 & 11 & 13 & 0.351 & \textbf{0.692} & \textbf{22.01$ \pm $8.40} & $0.90 \pm 0.03$\\
       & Semgrep & 15 & 24 & 10 & 25 & 0.081 & 0.462 & $27.09 \pm 0.52$ & \textbf{0.16$ \pm $0.01} \\
       & CodeQL & 23 & 29 & 5 & 17 & \textbf{0.428} & 0.676 & $65.96 \pm 0.15$ & $1.35 \pm 0.05$\\
     \midrule
     \multirow{3}{*}{securibench-micro.js}  
      & Joern & 92  & 38 & 6 & 25 & \textbf{0.650} & \textbf{0.856} & $58.90 \pm 3.31$ & $1.19 \pm 0.03$ \\
      & Semgrep & 5 & 43 & 1 & 112 & 0.020 & 0.081 & \textbf{34.12$ \pm $0.61} & \textbf{0.17$ \pm $0.01} \\
      & CodeQL & 85 & 31 & 13 & 32 & 0.431 & 0.791 & $134.47 \pm 1.54$ & $1.83 \pm 0.02$ \\
    \bottomrule
    \end{tabular}
    \label{tab:wholeresults}
\end{table}

\begin{figure}[h]
    \centering
    \includegraphics[width=0.9\linewidth]{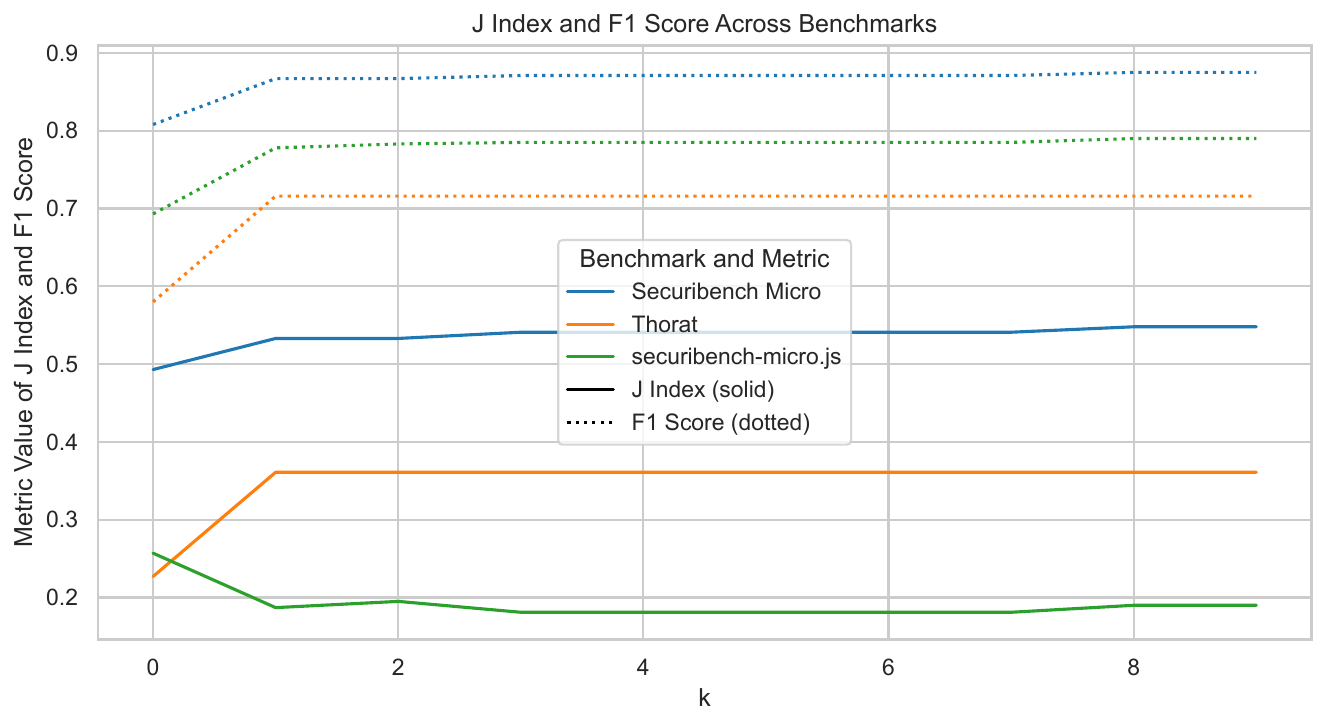}
    \caption{The results when exploring for an appropriate value for $k$ using each taint analysis benchmark.}
    \label{fig:taint_k_values}
    \Description{This figure describes a graph depicting the J index and F1 scores of the data-flow analysis when run on Securibench Micro, Thorat, and securibench-micro.js. A large variation in the J index and F score appears between $k \in [0, 3]$, followed by a slight increase in precision when $k = 8$ in Securibench Micro and securibench-micro.js. The early variation generally increases except for the J index on securibench-micro.js.}
\end{figure}

\begin{figure}[h]
    \centering
    \includegraphics[width=0.8\linewidth]{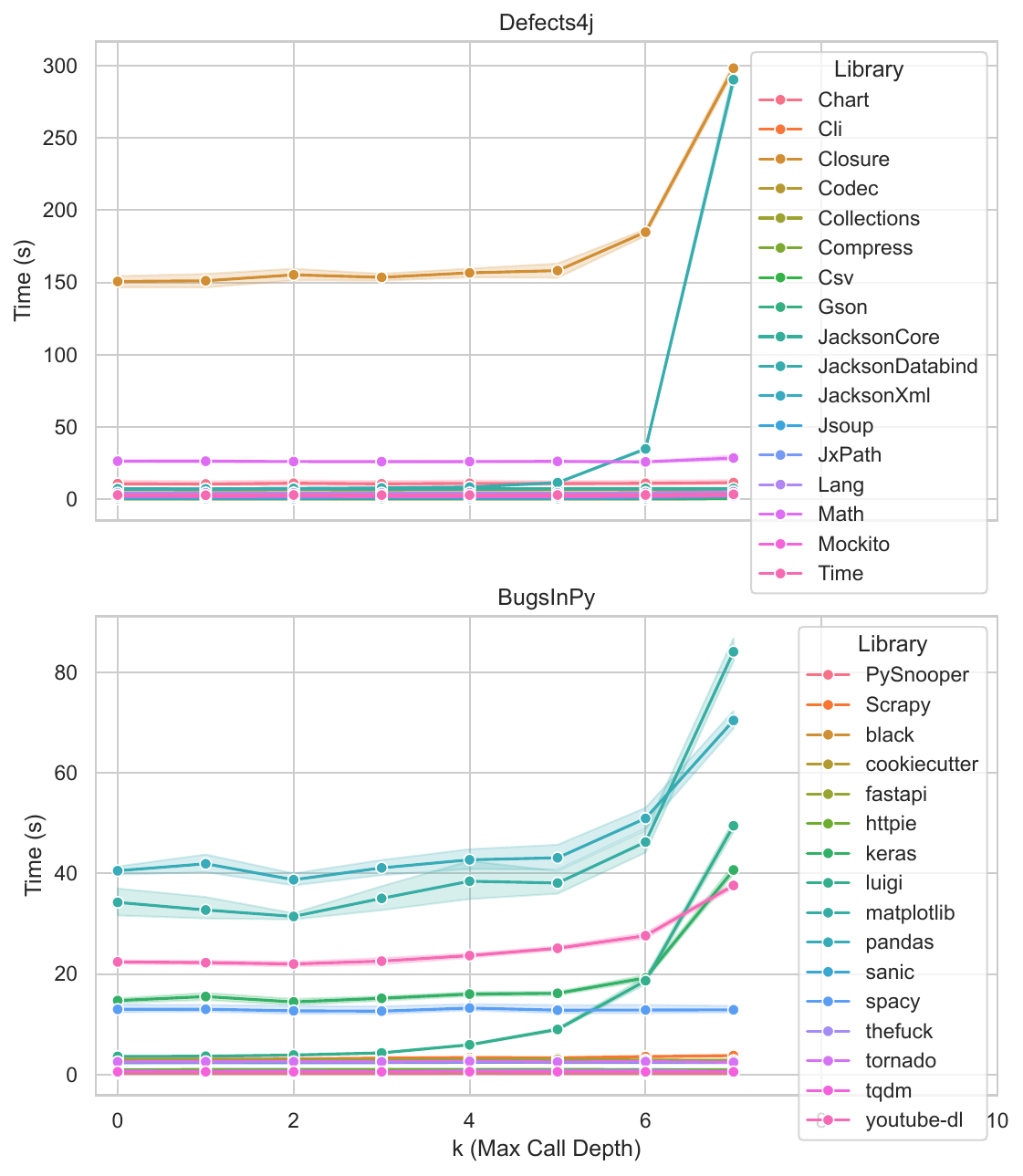}
    \caption{The performance of creating a code property graph and performing taint analysis on the programs of Defects4j and BugsInPy for varying values of $k \in [0, 7]$.}
    \label{fig:scalable_k_values}
    \Description{This figure describes two graphs depicting the performance of the data-flow analysis when run on the programs from Defects4j and BugsInPy. The performance is shown to plateau until, for a subset of programs, a state explosion begins at $k = 6$ where exponential time complexity is observed.}
\end{figure}

\end{document}